# Transfer Learning for sEMG-based Hand Gesture Classification using Deep Learning in a Master-Slave Architecture

Karush Suri*, Rinki Gupta
Electronics and Communication Engineering Department
Amity University Noida, Uttar Pradesh-201313, India
karushsuri@gmail.com, rgupta3@amity.edu

*Abstract*—Recent advancements in diagnostic learning and development of gesture-based human machine interfaces have driven surface electromyography (sEMG) towards significant importance. Analysis of hand gestures requires an accurate assessment of sEMG signals. The proposed work presents a novel sequential master-slave architecture consisting of deep neural networks (DNNs) for classification of signs from the Indian sign language using signals recorded from multiple sEMG channels. The performance of the master-slave network is augmented by leveraging additional synthetic feature data generated by long short term memory networks. Performance of the proposed network is compared to that of a conventional DNN prior to and after the addition of synthetic data. Up to 14% improvement is observed in the conventional DNN and up to 9% improvement in master-slave network on addition of synthetic data with an average accuracy value of 93.5% asserting the suitability of the proposed approach.

## I. INTRODUCTION

Study of hand gestures has gained significant importance in the past decade. With advancements in healthcare development and gesture recognition systems, signal analysis of hand gestures is extensively used for solving diagnostic learning problems and designing interactive human-machine interfaces. Sign language recognition, amputee treatment and kinematic study are some of the application scenarios of this diverse field [1,2]. The sEMG signals are a recording of the muscle potential caused by actuation of the muscle tendons when the limb is set in motion. Gestures performed by the limbs can broadly be assessed using three different sensing technologies namely vision-based, glove-based and using surface electromyogram (sEMG) signals [2]. The sEMG signals can be recorded in a non-invasive manner and the dry surface electrodes are reusable, making them convenient to use in assistive technology as well as consumer electronics.

Various hand gestures can be analyzed with the help of sEMG signals [2-3]. In sign language, the hand may be held in a particular configuration or posture, referred to as a static gesture, or it may be required to move the hand(s), which may be termed as a dynamic gesture. Distinction between various static gestures may require the use of an optimized classifier due to ample correlation between the sEMG signals recorded during static gestures. An artificial neural network (ANN) is used for such scenarios [4-5]. With a randomized initialization and operating over a number of iterations, the ANN provides suitable results on optimization of its topology [6]. Due to its adaptive nature, ANN can be used on a variety of data-sets including those containing sEMG signals [7-8].

Since ANNs do not have a definite topology, using a classifier with a dense architecture would be preferable [9]. This would ensure a further improvement in the learning process and absence of any additional functions [10]. A Deep Neural Network (DNN) consisting of multiple hidden layers is extensively used for this purpose. Additional hidden layers help in a more comprehensive evaluation of data by minimizing back-propagated errors [11]. Hidden layer computations are also essential for forward propagation as an increase in these layers would imply more activation, and thus more memory learning [12]. Augmentation of the performance of classifiers can further be achieved by making use of synthetically generated data. The synthetic data may be used for training and transferring similar characteristics to the learning stage [13]. Recursive Neural Networks (RNN) have been recently reported to be useful for generation of synthetic data samples. With their dynamic ability to make predictions as a result of previously processed data, these models can generate accurately similar sequences [14]. When a large number of gestures are to be classified for a given subject, the classification problem may become a cumbersome because sufficient number of observations of each gesture are required to be recorded for training a classifier, which would require a lot of time and effort. Another issue is that with a limited number of subjects, the training of the algorithm may remain incomplete leading to moderate accuracy.

In this work, first, the features extracted from the actual data are augmented with synthetically generated features modeling new subjects. A novel algorithm is proposed for the generation of synthetic data in the form of features corresponding to each subject by making use of Long Short-Term Memory (LSTM) cells trained on the original feature set. Section 2 describes sEMG gesture recording and the generation of additional synthetic data. Then, a novel sequential algorithm is proposed for classification of static and dynamic hand gestures by making use of DNNs, hereby referred to as the Master-Slave network, which is presented in Section 3. The master network distinguishes between the types of gestures, i.e., static gestures or dynamic gestures. Once the master network predicts the outputs, then the slave network classifies the specific gesture amongst the possible gestures in the identified type. The performance of the master-slave network is compared to the conventional DNN in Section 4. The performance of the DNNs are compared prior to and after the addition of synthetic data and Section 5 concludes the paper.

## II. GESTURE RECORDING AND GENERATION

### A. Data Corpus

The sEMG signals considered in this work are recorded from the surface of the skin over three muscles namely, Flexor Capri Ulnaris and Extensor Capri Radialis and Brachioradialis on the right arm of the subject. Fig. 1(a) shows the placement of sensors on the right arm of a subject. Fig. 1(b) provides complete experimental setup used in the recording of sEMG signals. The sEMG signals are recorded using the Delsys wireless EMG system. The sEMG signals are obtained with a sample rate of 1.1 kHz and a bit depth of 16 bits. The objective is to classify the sEMG signals from 10 gestures, which includes 5 static gestures and 5 dynamic gestures. The details of each gesture and its illustration are provided in Table I. The sEMG data has been collected from 4 healthy subjects in the age group of 22-30 years, all right-handed females. An audio stimulus of 3 seconds is played to the subject to indicate when the gesture is to be performed, with 5 seconds of rest in between each gesture. Each subject performed 20 repetitions for each gesture, the sEMG signals of which are recorded in continuation as one recording. A 2 min rest is given in between two recordings to avoid muscle fatigue.

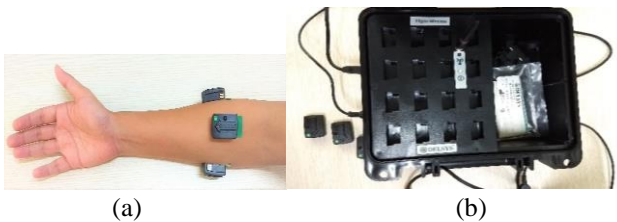

Fig. 1. Experimental Setup (a) Placement of Sensors, (b) sEMG Recording System

### B. Synthetic Feature Generation

Due to the limited number of subjects, there is a requirement for synthetically generating additional data to simulate more subjects. However, the synthetic data being added should be correlated to the originally recorded data for each gesture. Generating a complete signal from discrete samples is a time-consuming process. A more efficient way would be to obtain only the informative components of the signals. This is done by extracting features from the gestures of each subject. The most

TABLE I. GESTURES CONSIDERED FOR ANALYSIS

| Sign Type | Depiction of the Sign |
|---|---|
| Static | One, Two, Three, Four, Five |
| Dynamic | Sorry, Bold, Confident, Key, Win |

prominently used time-domain features are considered here [15-16]. These include:

*1) Integrated absolute value (IAV)*: Integrated absolute value (IAV) is the integration of all the absolute samples of the data. IAV of a random variable $x$ having $n$ values can be expressed mathematically as

$$\text{IAV} = \int_{i=1}^{n} |x_i|. \quad (1)$$

*2) Mean absolute value (MAV):* Mean absolute value (MAV) is the mean of absolute values of a segment of data. MAV of a random variable $x$ having $n$ values can be expressed mathematically as

$$\text{MAV} = \frac{1}{n}\sum_{i=1}^{n}|x_i|. \quad (2)$$

*3) Standard Deviation:* Standards deviation is defined as the square root of the variance

$$\sigma = \sqrt{\frac{1}{n-1}\sum_{i=0}^{n}(x_i - \mu)^2}, \quad (3)$$

where $\mu$ is the mean of the data, defined as $(\sum_{i=1}^{n} x_i)/n$.

*4) Root Mean Square (RMS):* The RMS value of a data segment is the root of the mean of squares of the data values. The RMS of a random variable $x$ having $n$ values can be expressed mathematically as

$$\text{RMS} = \sqrt{\frac{1}{n}\sum_{i=0}^{n}x_i^2}. \quad (4)$$

*5) Waveform Length (WL):* It is a cumulative variation of the sEMG that depicts the degree of variation about the signal, given as

$$\text{WL} = \sum_{k=1}^{n-1}|x_{k+1} - x_k|. \quad (5)$$

*6) Auto-Regressive Coefficients:* These are the constant coefficients ($a_r$) of the auto-regressive (AR) model $\hat{x}$ of the sampled instants of the sEMG signal, given as

$$\hat{x}_n = \sum_{k=1}^{p} a_k x_{n-k} + \varepsilon \quad (6)$$

where $p$ is the order of the AR model and $\varepsilon$ is the prediction error.

*7) Skewness:* The skewness of any signal segment with variable value $x$, having mean $\mu$ and standard deviation $\sigma$ is given as

$$\gamma = \frac{E\{x - \mu\}^3}{\sigma^3}, \quad (7)$$

where $E$ is the expectation operator.

*8) Mobility:* Mobility is defined as the ratio of the variance of the first derivative of the segment $x$ to the variance of the segment, given as

$$M = \frac{\nabla(var(x))}{var(x)}. \quad (8)$$

*9) Kurtosis:* The kurtosis of any signal segment with variable value $x$, having mean $\mu$ and standard deviation $\sigma$ is given as

$$\kappa = \frac{E\{x - \mu\}^4}{\sigma^4}. \quad (9)$$

The extracted features represent the cumulative correlation of a particular gesture varied from subject to subject enabling the modeling of a time-series data-set which can be evaluated with the help of a learning algorithm. An RNN consisting of multiple LSTM cells is suitable for evaluation of such a time series. In this work, the features are quantized to 20 levels corresponding to their range. Based on this quantization, the RNN treats the feature-set as a time series data with each repetition of a gesture being a time sample. The output is converted to a vector by virtue of one-hot encoding in order for it to be compatible with the LSTM cells. The LSTM cells help in back-propagating the errors through time and the network continues to iterate over a large number of time steps [17]. An LSTM cell consists of three gates, namely the input gate, the output gate and the forget gate. These gates shape an LSTM cell as an analogue storage element. Similar to a neural network node, the cells act upon the weights they receive. Since the weights tend to modulate layer by layer, these cells adapt and learn when to pass the information, block it and make predictions.

Once, the iterative process is complete and the input and output gates of all the cells are driven completely, the forget gate is activated. The gate clears the storage for avoiding any correlation between the processed time-series data and the data to be entered. Sequences in the processed time-series may then be used to offer predictions which when arranged in a consecutive order give rise to a new sequence corresponding to the same feature. De-quantization of the generated sequence is performed in order to obtain the feature vector. The process is then similarly repeated for all the features to generate the synthetic data-set

### III. CLASSIFICATION OF HAND GESTURES

In this work, classification of ten hand gestures is considered. As depicted in Fig. 2(a), classification is carried out by making use of the proposed master-slave network and its performance is compared to a conventional DNN. The proposed approach makes use of DNNs which consist of multiple hidden layers capable of learning the intricate aspects of data. In the master-slave network, learning takes place in two phases, forward propagation followed by back-propagation. In the forward propagation phase, the DNN takes in the features of each gesture as inputs and carries out computations at each node by virtue of an activation function, the sigmoid function in this case. These computations iteratively update the value of cost function, which is used to optimize the network. The activation function and cost function can be mathematically expressed as

$$g(x^{(i)}) = \frac{1}{1 + e^{-x^{(i)}}}. \quad (10)$$

and

$$J(\Theta) = -\frac{1}{m}\Big[\sum_{i=1}^{m}\sum_{k=1}^{K}\big\{y_k^{(i)}log(h_\Theta(x^{(i)}))_k + (1 - y_k^{(i)})(1 - log(h_\Theta(x^{(i)})))_k\big\}\Big]. \quad (11)$$

In (11), '*m*' is the total number of signals to be classified, '*K*' is the total number of classes, '*L*' is the total number of layers, '$h_\Theta$' is the hypothesis yielding the prediction '*y(i)*' corresponding to input '*x(i)*'.

Fig. 2(b) depicts the sequential architecture of the master-slave network. The network is composed of two multi-layer perceptrons arranged in a sequential order. Each DNN in the architecture consists of input nodes being equal to the number of features and the number of hidden nodes in each hidden layer being 1.5 times the number of features. For the sake of reducing complexity and computational expense, the number of hidden layers has been empirically set to 4.

The network operates in reverse direction in order to compute the errors '$\delta(j)$' at each neuron '*j*'. Computation of errors begins from the last hidden layer by simply calculating the difference between the output of the layer given as '$a_j(5)$' and the predicted output '$y_j$'. Mathematical expressions for errors at neurons of last hidden layer and at neurons of other layers respectively are represented as

$$\delta_j^{(5)} = a_j^{(5)} - y_j. \quad (12)$$

$$\delta_j^{(L)} = a_j^{(L)} * (1 - a_j^{(L)}). \quad (13)$$

For each iteration in DNN, the input undergoes feed-forward propagation and back-propagation. The back-propagated errors are used to update the gradient values used during the process of optimization. For each input to the layer '*j*' computation of the '$\Delta_{ij}$' value results in the new gradient. This can be expressed as

$$\Delta_{ij} := \Delta_{ij} + a_j^{(z)} * \delta_j^{(z)} \; ; z \in [1,5] \quad (14)$$

$$D_{ij} = \frac{1}{m}\Delta_{ij} + a_j * \Theta_{ij}^{(z)} \; ; z \in [1,5] \quad (15)$$

$$\frac{d\,J(\Theta)}{d\Theta_{ij}^{(z)}} = D_{ij} \; ; z \in [1,5] \quad (16)$$

The process is terminated when the cost function achieves the global minimum, i.e. when the network is optimized. Predicted values obtained corresponding to each signal '*m*' are used to compute the classification accuracy.

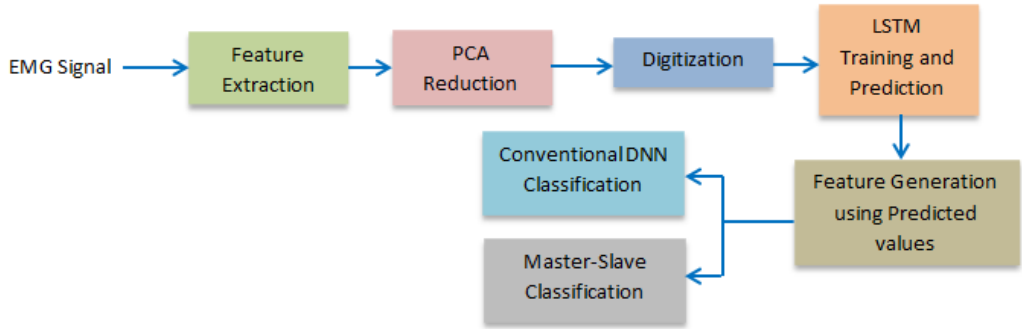

(a) Proposed master-slave network vs. conventional DNN

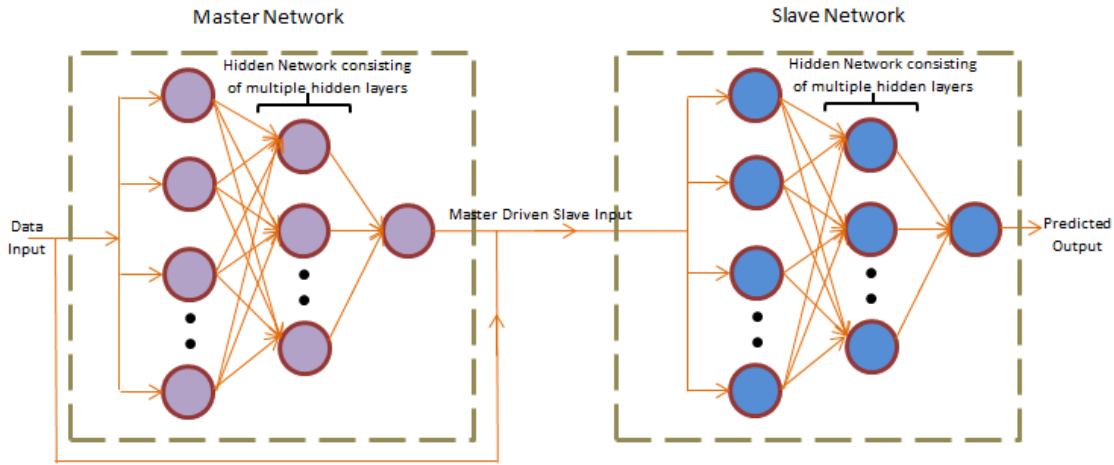

(b) Proposed sequential architecture of the master-slave network

Fig. 2. Comparison of Master-Slave operation with conventional DNNs

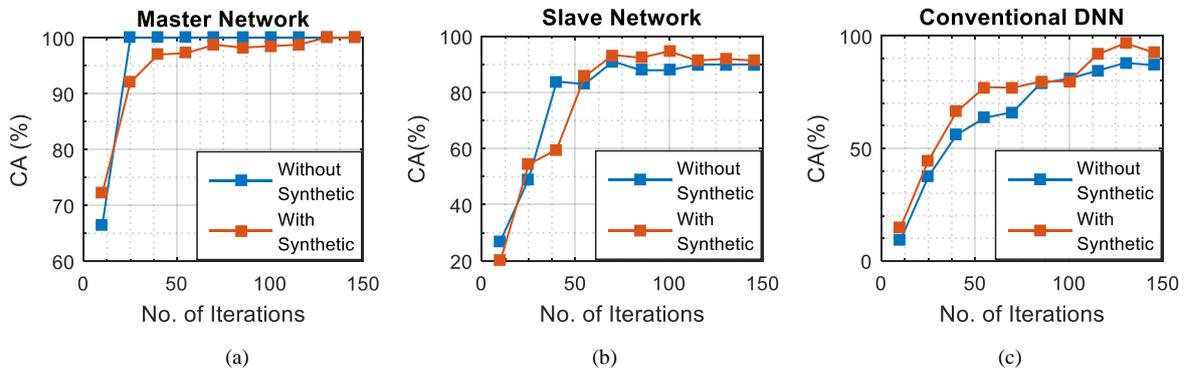

Fig. 3. Learning curves for subject 3 over 150 iterations (a) Master network, (b) Slave network, (c) Conventional DNN

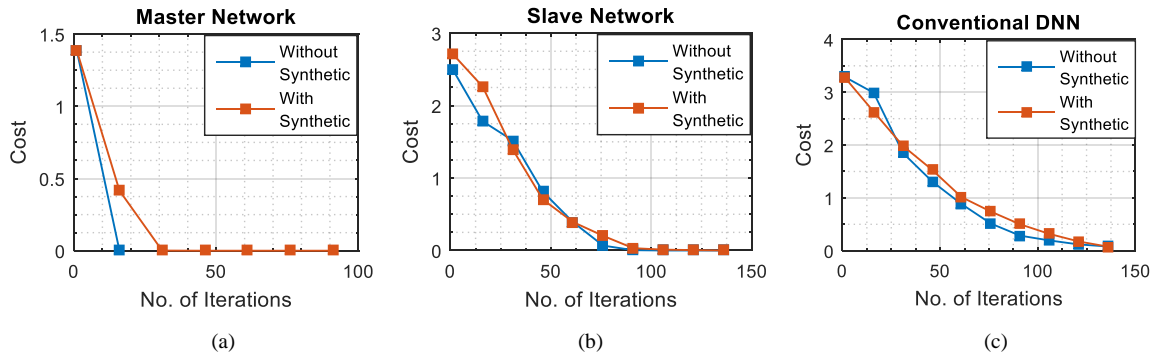

Fig. 4. Cost convergence for subject 3 over 150 iterations (a) Master network, (b) Slave network, (c) Conventional DNN

TABLE II. CA VALUES FOR ALL THE SUBJECTS PRIOR TO AND AFTER THE ADDITION OF SYNTHETIC DATA (OBTAINED OVER 150 ITERATIONS)

| Subject No. | Master Network | | Slave Network | | Conventional DNN | |
| --- | --- | --- | --- | --- | --- | --- |
| | Without Synthetic Data | With Synthetic Data | Without Synthetic Data | With Synthetic Data | Without Synthetic Data | With Synthetic Data |
| Subject-1 | 100% | 96.757% | 92.008% | **93%** | 88.482% | **95.997%** |
| Subject-2 | 98.507% | 99.25% | 91.979% | **97%** | 96.495% | **98.496%** |
| Subject-3 | 99.502% | 98.251% | 89.008% | **93.33%** | 85.512% | **95.01%** |
| Subject-4 | 99.495% | 98.008% | 81.075% | **90.667%** | 71.559% | **85.265%** |

## IV. RESULTS AND DISCUSSION

Addition of synthetic data results in an improvement in the classification accuracy (CA) of the master-slave network and the conventional DNN. Performance of the classification models is augmented by making use of the synthetic features during the training process. Fig. 3 depicts the learning curves of networks where CA is plotted with respect to the number of iterations. As the number of iterations increases, CA increases non-linearly indicating the learning stage of the models. The variations in CA for a given model are compared in the absence and presence of synthetic data. Final CA values of the models are greater in the presence of synthetic data when compared to the CA values in the absence of synthetic data, hence validating the efficacy of the proposed approach. CA approaches 100% in the case of master network at the end of the training phase and improves from 89.93 % to 91.47% in the case of slave network and from 87% to 92.37% in the case of conventional DNN on completion of the training phase.

The performance of the models may also be assessed by comparing their convergence with and without the synthetic data. Fig. 4 presents the cost convergence of the networks varied against the number of iterations. At the end of each iteration, the value of cost progresses and reaches a minimum value. At the minimum cost value, the model is said to have converged and cannot learn any significant characteristics beyond this point. Cost of the master network converges to zero, as shown in Fig. 4a. In the case of slave network also (Fig. 4b), the cost converges to zero, but it is 0.064 in the case of conventional DNN, as shown in Fig. 4c. Convergence, in the case of master network, takes place at a faster rate due to the existence of only two classes used during the learning stage. On addition of synthetic data, convergence of the models takes place in exactly the same number of iterations, thus preventing any extra computational expense and providing better CA values with the same convergence.

Improvement in CA for the master-slave network and conventional DNN can be obtained for all the subjects by making use of synthetic data. Table II enlists CA values obtained by using master-slave network and conventional DNN, with and without the addition of synthetic data. A peak improvement in the master-slave network is observed to be from 81.075% to 90.667% in the case of subject no. 4. In the case of conventional DNN, peak CA value improves from 71.559% to 85.265% in the case of subject no. 4 as well. Approximately 1% to 9% improvement in CA values is observed for master-slave network and 2% to 14% improvement for conventional DNN. The enhanced performance of the two models when synthetic data is used, depicts the suitability of the synthetic data approach.

## V. Conclusion

In order to provide an optimized algorithm capable of distinguishing between hand gestures in the Indian Sign Language from the sEMG signals, sufficient data is required for training a classifier. In this work, static hand gestures recorded from wearable sEMG sensors are classified with the help of a novel sequential master-slave architecture of DNNs. The performance of the network is analysed with regard to a conventional DNN. Further augmentation of classification process is carried out by leveraging synthetic data generated in the form of features by LSTM-RNN. Classification accuracy values improve from 1%-9% in the case of master-slave network and 2%-14% in the case of conventional DNN. Increase in classification accuracy validates the suitability of the proposed algorithm.

## Acknowledgment

The authors would like to recognize the funding support provided by the Science & Engineering Research Board, a statutory body of the Department of Science & Technology (DST), Government of India, SERB file number ECR/2016/000637.

## References

[1] S. Cobos, M. Ferre, "Human hand descriptions and gesture recognition for object manipulation", Computer Methods in Biomechanics and Biomedical Engineering, Vol. 13, No. 3, pp. 305-317, 2010.

[2] J. Wu, L. Sun, R. Jafari, "A Wearable System for Recognizing American Sign Language in Real-Time Using IMU and Surface EMG Sensors," IEEE J. Biomedical and Health Informatics, Vol. 20, No. 5, pp. 1281-1290, 2016.

[3] Christos Sapsanis, George Georgoulas, Anthony Tzes, and Dimitrios Lymberopoulos, "Improving EMG based Classification of basic hand movements using EMD", 35th Annual International Conference of the EMBS, pp. 5754-5757, 2013.

[4] Md. R. Ahsan, M. I. Ibrahimy, O. O. Khalifa, "The use of Artificial Neural Network in the Classification of EMG Signals", Third FTRA International Conference on Mobile, Ubiquitous, and Intelligent Computing, pp. 225-229, 2012.

[5] A. M. Khan, Y. K. Lee, T. S. Kim, "Accelerometer Signal-based Human Activity Recognition Using Augmented Autoregressive Model Coefficients and Artificial Neural Nets", 30th Annual International IEEE EMBS Conference, 2008.

[6] S. H. Roy, M. S. Cheng, S. S. Chang, J. Moore, G. De Luca, S. H. Nawab, C. J. De Luca, "A Combined sEMG and Accelerometer System for Monitoring Functional Activity in Stroke", IEEE Transactions on Neural Systems and Rehabilitation Engineering, Vol. 17, No. 6, pp. 585-594, 2009.

[7] Naser El-Sheimy, Kai-Wei Chiang, and Aboelmagd Noureldin, "The Utilization of Artificial Neural Networks for Multisensor System Integration in Navigation and Positioning Instruments", IEEE Transactions on Instrumentation and Measurement, Vol. 55, NO. 5, pp. 1606-1615, 2006.

[8] P. Shukla, I. Basu, D. Graupe, D. Tuninetti, K. V. Slavin," A Neural Network-based Design of an on-off Adaptive Control for Deep Brain Stimulation in Movement Disorders", 34th Annual International Conference of the IEEE EMBS, 2012.

[9] P. Różycki, J. Kolbusz and B.M. Wilamowski, Rzeszów, "Dedicated Deep Neural Network Architectures and Methods for Their Training", IEEE 19th International Conference on Intelligent Engineering Systems, pp. 1-5, 2015.

[10] Michael Wand and Tanja Schultz, "Pattern Learning with Deep Neural Networks in EMG-based Speech Recognition", pp. 1-5, 36th Annual Conference of the IEEE EMBC, 2014.

[11] Wei-Long Zheng, Bao-Liang Lu, "Investigating Critical Frequency Bands and Channels for EEG-Based Emotion Recognition with Deep Neural Networks", IEEE Transactions on Autonomous Mental Development, Vol. 7, No. 3, 2015.

[12] Michael L. Seltzer, Dong Yu, Yongqiang Wang, "An Investigation of Deep Neural Networks for Noise Robust Speech Recognition", IEEE International Conference on Acoustics, Speech and Signal Processing, pp. 1-5. 2013.

[13] Ulysse Cotˆe-Allard, Cheikh Latyr Fall, Alexandre Drouin, Alexandre Campeau-Lecours, Clement Gosselin, Kyrre Glette, Franc¸ois Laviolette, and Benoit Gosseli, "Deep Learning for Electromyographic Hand Gesture Signal Classification by Leveraging Transfer Learning", ArXiv, pp. 1-13, 2018.

[14] Abdulmajid Murad and Jae-Young Pyun, 6 November 2017, "Deep Recurrent Neural Networks for Human Activity Recognition", Sensors, Vol. 17, 2556, pp. 1-17.

[15] Rinki Gupta, Ankita Kulshreshtha, "Analysis of dual channel surface electromyogram using second order and higher order spectral features", IEEE 2nd International Conference on Communication Control and Intelligent Systems, pp. 1-6, Nov. 2016.

[16] Angkoon Phinyomark, Pornchai Phukpattaranont, Chusak Limsakul, 2012, "Feature reduction and selection for EMG signal classification", Expert Systems with Applications, vol. 39, pp. 7420–7431, 2012.

[17] Intan Nurma Yulitaa, Mohamad Ivan Fananya, Aniati Murni Arymuthy, October 2017, "Bi-directional Long Short-Term Memory using Quantized data of Deep Belief Networks for Sleep Stage Classification", 2nd International Conference on Computer Science and Computational Intelligence, ICCSCI, Elsevier.